\documentclass[conference]{IEEEtran}

\IEEEoverridecommandlockouts

\usepackage{cite}
\usepackage{amsmath,amssymb,amsfonts}
\usepackage{algorithmic}
\usepackage{graphicx}

\usepackage{textcomp}
\usepackage{xcolor}
\usepackage{CJKutf8}
\usepackage{tabularray}
\usepackage{subfigure}
\usepackage{caption}

\def\BibTeX{{\rm B\kern-.05em{\sc i\kern-.025em b}\kern-.08em
    T\kern-.1667em\lower.7ex\hbox{E}\kern-.125emX}}
\begin{document}

\columnsep 0.241in
\topskip 0.19in

\begin{CJK}{UTF8}{gbsn}
\title{Deep Joint Source-Channel Coding for Wireless Image Transmission with Entropy-Aware Adaptive Rate Control}

\author{\IEEEauthorblockN{Weixuan Chen$^{*}$, Yuhao Chen$^{\dag}$, Qianqian Yang$^{*}$\textsuperscript{\textsection}, Chongwen Huang$^{*}$, Qian Wang$^{\ddag}$, Zhaoyang Zhang$^{*}$}
\IEEEauthorblockA{{
}
{$^{*}$ College of Information Science and Electronic Engineering, Zhejiang University, Hangzhou 310007, China}\\
{$^{\dag}$ College of Control Science and Engineering, Zhejiang University, Hangzhou 310007, China}\\
{$^{\ddag}$ College of Information Engineering, Zhejiang University of Technology, Hangzhou 310023, China}\\
\{12231075, csechenyh, \textsuperscript{\textsection}qianqianyang20, chongwenhuang, ning\_ming\}@zju.edu.cn \\
  {wangqian18@zjut.edu.cn}
}
}


\maketitle

\begin{abstract}
Adaptive rate control for deep joint source and channel coding (JSCC) is considered as an effective approach to transmit sufficient information in scenarios with limited communication resources.
We propose a deep JSCC scheme for wireless image transmission with entropy-aware adaptive rate control, using a single deep neural network to support multiple rates and automatically adjust the rate based on the feature maps of the input image and their entropy, as well as the channel conditions. 
In particular, we maximize the entropy of the feature maps to increase the average information carried by each transmitted symbol during the training.
We further decide which feature maps should be activated based on their entropy, which improves the efficiency of the transmitted symbols.
We also propose a pruning module to remove less important pixels in the activated feature maps in order to further improve transmission efficiency. 
The experimental results demonstrate that our proposed scheme learns an effective rate control strategy that reduces the required channel bandwidth while preserving the quality of the reconstructed images. 
\end{abstract}


\section{Introduction}
Conventional wireless image transmission systems typically perform source coding and channel coding of the source image separately. 
Source coding first compresses the source image to remove redundant information.
Then, channel coding adds redundant bits to the source-coded image to ensure reliable wireless transmission.
According to Shannon's separation theorem\cite{cover1999elements}, this structure requires the transmission of infinitely long codewords to achieve theoretical optimality.
However, in practical scenarios such as autonomous driving and telemedicine, long code block lengths are difficult to achieve.
Therefore, separation-based coding schemes may not be optimal for wireless transmission.

To address this issue, deep learning techniques have been introduced to wireless image transmission systems to offer significant performance improvements. 
Bourtsoulatze et al. proposed the joint source and channel coding (JSCC) algorithm for wireless image transmission based on deep learning\cite{bourtsoulatze2019deep}, which is referred to as the deep JSCC scheme. 
In this scheme, the authors train a deep learning model to extract a number of feature maps from the source image, with the non-trainable communication channel incorporated into the training process. 
Experimental results have shown that the deep JSCC scheme achieves better performance than the separation-based digital transmission scheme. 
Based on the deep JSCC scheme, several studies have been conducted to further improve its performance. 
Kurka et al.\cite{kurka2020deep} proposed DeepJSCC-f, which introduces the channel output feedback to the deep JSCC scheme. This achieves variable-length transmission and reduces the required channel bandwidth. 
Zhang et al.\cite{zhang2022wireless} proposed MLSC-image, a multi-level semantic communication system for image transmission. It extracts both high-level and low-level image semantic features to improve image reconstruction performance.

However, there are still some aspects of the pioneering deep JSCC scheme that could be improved. 
On one hand, deep JSCC maps the pixel values of the input image to the complex-valued channel input symbols, treating each symbol equally with the same channel resources. 
However, for different transmission purpose, the transmitted symbols may have different importance. 
On the other hand, the transmission rate of the deep JSCC scheme is fixed, which limits the transmission of complex image content when communication resources are limited. 
To address this limitation, the deep JSCC scheme could be made more flexible and effective by dynamically adjusting the transmission rate to transmit important information with limited channel bandwidth.

Regarding the first aspect, we believe that symbols from feature maps with higher entropy are more important because they have more information on average and thus may contribute to better image reconstruction.
As for the second aspect, several studies on multi-rate deep JSCC schemes have been conducted. 
Kurka et al.\cite{kurka2019successive}\cite{kurka2021bandwidth} studied the problem of adaptive bandwidth image transmission over wireless channels. 
Yang et al.\cite{yang2022deep} supported multiple rates using a single deep neural network that can dynamically adjust its rate based on the channel signal-to-noise ratio (SNR) and the feature maps. 
However, these studies only considered reducing the number of feature maps to be transmitted, ignoring the fact that the length of each feature map can also be reduced.
In contrast, Zhou et al. \cite{zhou2022adaptive} proposed a multi-bit length semantic encoding scheme that can reasonably decrease the length of transmitted information bits and minimize the overhead required for the correct delivery of information. 
Inspired by \cite{zhou2022adaptive}, we believe that not only some feature maps may be unimportant, but even some of the pixels within an important feature map are also unimportant.

In this paper, we propose an entropy-aware deep JSCC scheme that can automatically adjust its rate based on the feature maps and their entropy, as well as the channel conditions. 
Specifically, we define the feature maps that need to be transmitted as the \textbf{activated feature maps}. 
We select the activated feature maps from all feature maps of the image and then prune them to increase the flexibility of the transmission rate. 
The main contributions of this paper are as follows:

$\bullet$ \textbf{Entropy-Aware}: We maximize the entropy of the feature maps during training to increase the average amount of information carried by each transmitted symbol and determine the activated feature maps based on their entropy. This benefits the image reconstruction performance of the model.

$\bullet$ \textbf{Sparse Feature Maps}: We prune the activated feature maps while maintaining their structural information. This reduces the transmission of unimportant symbols, resulting in reduced channel bandwidth usage.

$\bullet$ \textbf{Automatic Rate Adaption}: The selection and pruning of feature maps are fully automatic. We use two policy networks to determine the activated feature maps and the pruning ratio for these activated feature maps, respectively.

Our experimental results demonstrate that the proposed method is effective in selecting the activated feature maps and pruning them under different channel conditions.
Compared to existing methods, our proposed method achieves a notable improvement in image reconstruction performance, with a PSNR increase of approximately 0.11-0.57dB. 

\section{Our Proposed Deep JSCC Scheme with Entropy-Aware Adaptive Rate Control}

\subsection{Overall Structure}
In this section, we propose a deep JSCC scheme that can automatically adjust its rate based on the feature maps and their entropy, as well as the channel SNR, while maintaining high image reconstruction performance. 
The proposed scheme consists of several components, namely a semantic encoder, a feature map selection and pruning module, a channel encoder, a communication channel, a channel decoder, and a semantic decoder. 

\begin{figure*}[htbp]

\begin{center}
\centerline{\includegraphics[width=1\linewidth]{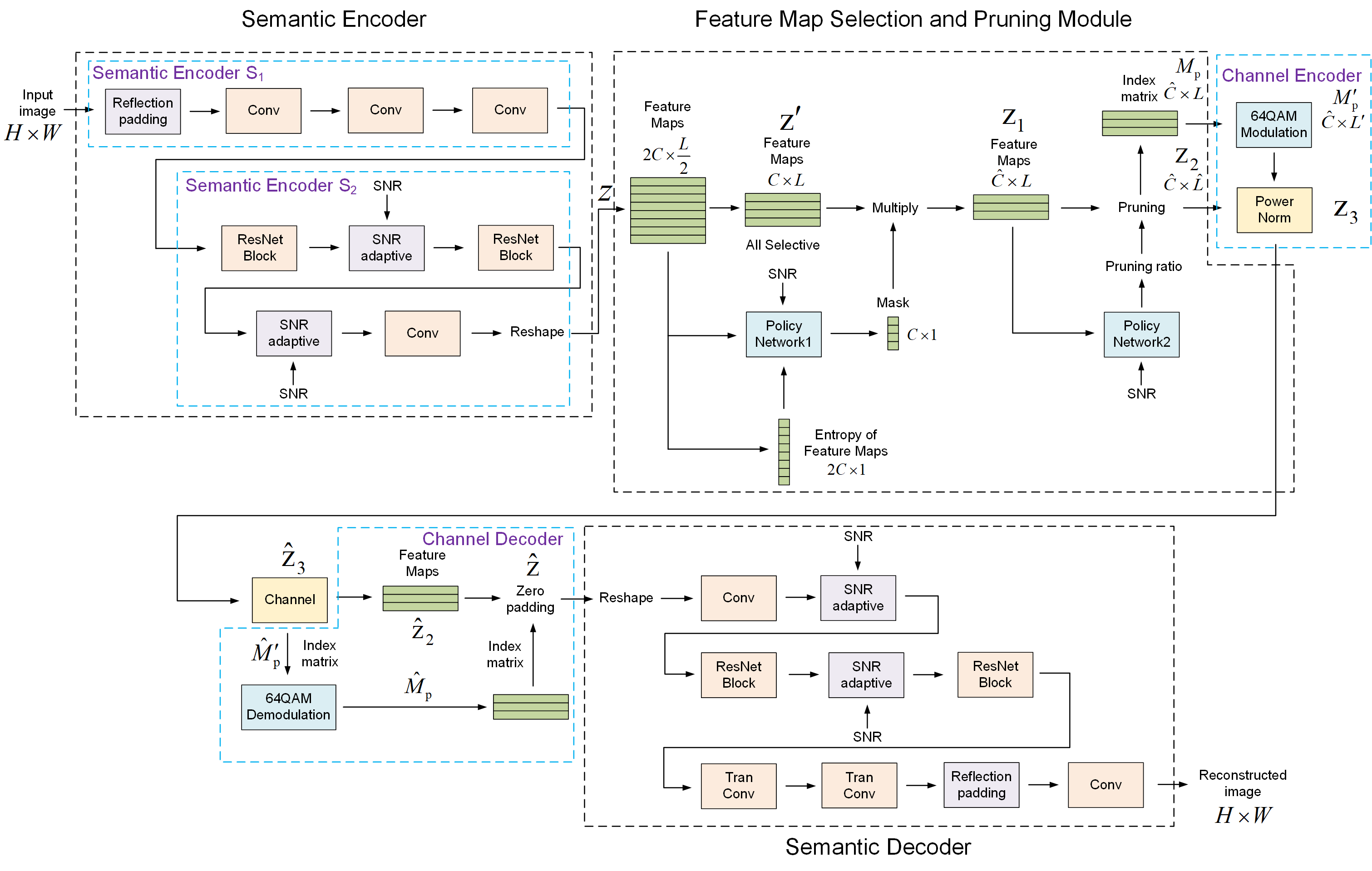}}
\caption{The overall structure of our proposed deep JSCC scheme.}
\label{overall_architecture}
\end{center}
\vskip -0.3in
\end{figure*}

To transmit an image $x$, we feed it into the semantic encoder for extracting feature map $z = f_{\mathrm{se}}(x)$.
Then in the feature map selection and pruning module, both $z$ and its entropy $H_{z}$ are provided as inputs to a policy network $P_{1}$ in order to generate a mask $M$, based on which the activated feature map $z_{1}$ is selected. 
$z_{1}$ is then input to another policy network $P_{2}$ to obtain its pruning ratio, which is applied by the feature map pruning module to prune $z_{1}$, resulting in the pruned feature map $z_{2}$ and the pruning index matrix $M_\mathrm{p}$. 
After obtaining $z_{2}$ and $M_\mathrm{p}$, they are fed into the channel encoder to yield a complex-valued transmitted signal, represented as $z_{3}$. $z_{3}$ is passed through the communication channel and becomes $\hat{z_{3}}$.
$\hat{z_{3}}$ contains two components: the received signal $\hat{z_{2}}$ and the received modulated pruning index matrix $\hat{M_\mathrm{p}'}$. 
At the receiver, the channel decoder maps $\hat{z_{2}}$ back to $\hat{z}$ according to $\hat{M_\mathrm{p}'}$, from which the semantic decoder then reconstructs the image $\hat{x}$. 
A comprehensive illustration of our proposed deep JSCC scheme is presented in Fig.~\ref{overall_architecture}. Next, we will introduce each module in detail. 

\subsection{Semantic Encoder and Decoder}
The semantic encoder takes $x$ as input and outputs the feature map $z$, while the semantic decoder reconstructs the image from $\hat{z}$. 
The semantic encoder consists of two components, $S_{1}$ and $S_{2}$.
The semantic encoder $S_{1}$ consists of three convolutional layers, while semantic encoder $S_{2}$ includes two ResNet blocks, two SNR adaptive modules, and one convolutional layer. The structure of the SNR adaptive module is identical to that in \cite{yang2022deep}.
The structure of the semantic decoder is in reverse order to that of the semantic encoder. We also note that some convolutional layers in the semantic encoder become transpose convolutional layers for upsampling in the semantic decoder.


\subsection{Feature Map Selection and Pruning Module}
The feature map selection and pruning module selects the activated feature maps and prunes them. 
Firstly, we define the dimension of $z$ as $2C \times \frac{L}{2}$, where $2C$ represents the number of feature maps and $\frac{L}{2}$ is the length of each feature map. 
We calculate the entropy of each feature map $z_{i}$ as $H(z_{i})$, where $i$ is the order index.
Then, in order to increase the difference between the entropy values of the feature maps, we apply the softmax function
	$H'\left( z_{i} \right) = \frac{e^{H(z_{i})}}{\sum\limits_{i = 1}^{2C}e^{H(z_{i})}}$
to normalize $H(z_{i})$. 
The overall entropy matrix $H_{z}$ is obtained by concatenating all $H'(z_{i})$.
Then every two feature maps in $z$ are concatenated to form a concatenated feature map $z' \in \mathbb{R}^{C \times L}$, since the feature maps are input to the communication channel with one half as the real part and the other half as the imaginary part. 
After concatenation, all $C$ feature maps can be selected as activated feature maps since this can further increase the flexibility of our transmission strategy. However, only some of the feature maps were selective in the previous method \cite{yang2022deep}. 

\begin{figure}[t]
\vskip -0.1in
\begin{center}
\centerline{\includegraphics[width=0.9\linewidth]{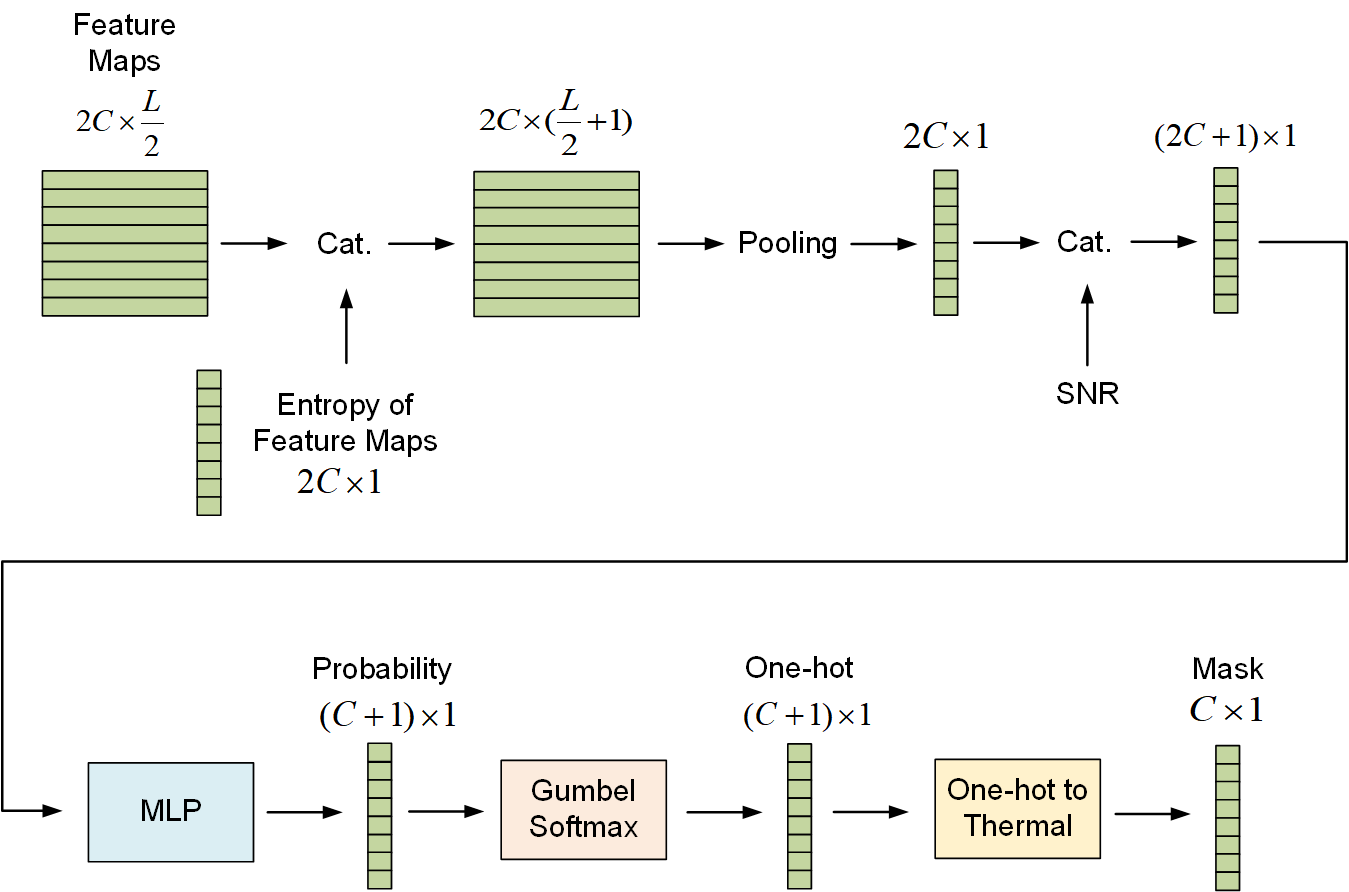}}
\caption{The structure of policy network $P_{1}$.}
\label{policy1}
\end{center}
\vskip -0.3in
\end{figure}

Next, we feed $z$ and $H_{z}$ into a policy network $P_{1}$. This network outputs a binary mask $M \in \left\{ 0,1 \right\}^{C \times 1}$ that indicates which feature maps should be activated. 
$P_{1}$ learns to select the activated feature maps based on all the feature maps and their entropy, as well as the channel SNR, whose structure is shown in Fig.~\ref{policy1}. 
As shown in Fig.~\ref{policy1}, we first concatenate $z$ and $H_{z}$ to construct an information matrix for the feature maps. We then average each row of this matrix and concatenate the resulting matrix with the channel SNR. This generates a $(2C + 1) \times 1$ matrix.
The probabilities for each possible selection are generated by passing this matrix into a two-layer multi-layer perceptron (MLP).
To overcome the non-differentiability problem caused by the discrete nature of the sampling process, we utilize Gumbel-Softmax to sample the selection as a one-hot vector. This vector is then transformed into the mask $M$.
The adaptive transmission mask has a length of $C$, where each ``1" in the mask means that the feature map at the corresponding position in $z'$ will be activated, and vice versa.
We then multiply the concatenated feature map $z'$ by $M$ to obtain the activated feature map $z_{1}$. 
Note that the inputs of $P_1$ are $z$ and $H_{z}$, while the activated feature maps are selected from $z'$. 
At the receiver, feature maps that are not activated are zero-padded. We denote the total number of activated feature maps as $\hat{C} = {\sum\limits_{i = 1}^{C}M_{i}}$. 

\begin{figure}[t]
\vskip -0.1in
\begin{center}
\centerline{\includegraphics[width=0.8\linewidth]{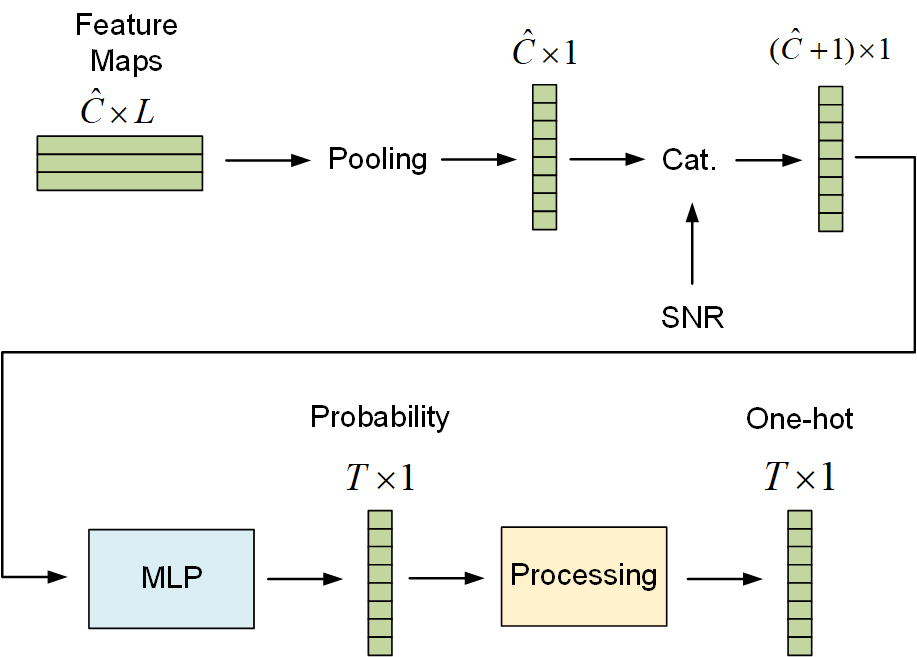}}
\caption{The structure of policy network $P_{2}$.}
\label{policy2}
\end{center}
\vskip -0.3in
\end{figure}

In the next step, we use another policy network $P_{2}$ to determine the pruning ratio for the activated feature maps. $P_{2}$ takes $z_{1}$ as its input and outputs a one-hot vector $V$ of length $T$, where the $T$ denotes the number of possible pruning ratios. Its structure is shown in Fig.~\ref{policy2}.
As shown in Fig.~\ref{policy2}, we first average each row of $z_{1}$ and concatenate the resulting matrix with the channel SNR. The concatenated matrix is then fed into the MLP, which consists of two fully connected layers and a softmax function at the end. By utilizing the maximum probability value, we generate the one-hot vector of length $T$. In this paper, we set $T$ to 5 and define the possible pruning ratios as $a \in \left\{ 0,0.2,0.25,0.3,0.35 \right\}$. 
Once we obtain the one-hot vector $V$, we apply pruning to $z_{1}$ based on $V$ to obtain the pruned feature map $z_{2}$. The dimension of $z_{2}$ is $\hat{C} \times \hat{L}$. Specifically, inspired by the commonly used $l$1-norm based pruning rule in model pruning\cite{han2015learning}, for each activated feature map, we regard its pixel with smaller $l$1-norm (i.e., pixels with smaller absolute numerical values) as being less important. 
During the pruning process, we remove a portion of pixels with relatively small $l$1-norm in each activated feature map according to the pruning ratio obtained from $V$.
However, the positions of these removed pixels in the activated feature maps contain important structural information of the input image.
Therefore, we record the position indices of the pruned pixels in a pruning index matrix $M_\mathrm{p} \in \left\{ 0,1 \right\}^{\hat{C} \times L}$, where the ``1'' indicates that the pixel at that position has been pruned.
It is essential to transmit $M_\mathrm{p}$ to the receiver so that the receiver can zero-pad the pruned pixels according to $M_\mathrm{p}$, thus recover the structure of the activated feature maps.

\subsection{Channel Encoder and Decoder}
In the channel encoder, we use 64-quadrature amplitude modulation (64-QAM) to modulate $M_\mathrm{p}$, resulting in a length of each vector of $L' = \left\lceil \frac{L}{6} \right\rceil$. 
Then, we obtain the modulated pruning index matrix $M_\mathrm{p}'$ for the activated feature maps. It has a dimension of $\hat{C} \times L'$ pixels. 
We input the pruned feature map $z_{2}$ and the modulated pruning index matrix $M_\mathrm{p}'$ into the power normalization module, which generates a complex-valued transmitted signal $z_{3} \in \mathbb{C}^{\hat{C} \times {(\frac{\hat{L} + L'}{2})}}$ with unit average power.
After that, we input $z_{3}$ into the noisy wireless channel and output $\hat{z_{3}}$. $\hat{z_{3}}$ contains the received signal $\hat{z_{2}}$ and the received modulated pruning index matrix $\hat{M_\mathrm{p}'}$.
Finally, we feed $\hat{z_{2}}$ and $\hat{M_\mathrm{p}'}$ into the channel decoder.

In the channel decoder, we demodulate $\hat{M_\mathrm{p}'}$ and zero-pad the pruned pixels according to $\hat{M_\mathrm{p}}$ to recover the structure of the activated feature maps. Subsequently, we zero-pad the feature maps that are not activated. The output of the channel decoder is $\hat{z}$.

\subsection{Communication Channel}
The communication channel is modeled as a non-trainable layer that adds random perturbation to the transmitted symbols.
In this paper, we consider AWGN wireless channel.
The output of the channel can be expressed as $\hat{z_{3}} = z_{3} + n$, where $n$ represents the additive Gaussian noise.  
The channel SNR, which determines the level of noise in the channel, is known by both the transmitter and receiver, allowing the semantic encoder, semantic decoder, $P_{1}$, and $P_{2}$ to adapt to different channel conditions.


\subsection{Transmission Rate}
Our objective is to maintain the quality of reconstructed image while reducing the channel bandwidth usage.
We measure the transmission rate based on the wireless channel usage per pixel (CPP).
If the input image has a dimension of $3 \times H \times W$ pixels, then 
	$\mathrm{CPP} = \frac{\hat{C}\left( \hat{L} + L' \right)}{2HW}$.
Recall that $\hat{C}$ represents the number of activated feature maps, $\hat{L}$ represents the length of the pruned activated feature maps, and $L'$ represents the number of columns of the modulated pruning index matrix $M_\mathrm{p}'$. When no feature maps are pruned, $L^{'} = 0$.
Hence, the value of CPP mainly depends on the number and length of the activated feature maps, as well as the number of pixels removed from the activated feature maps.



\subsection{Loss Function and Training Strategy}
To train the proposed deep JSCC scheme, we incorporate three terms into our loss function $\mathcal{L}$, which is defined as :
\begin{equation}
\label{eq1}
\mathcal{L} = \mathbb{E}\left\lbrack {\left\| {x - \hat{x}} \right\|_{2}^{2} + \alpha\frac{\hat{L}}{L}{\sum\limits_{i = 1}^{C}M_{i}} - \beta\mathbb{E}\left( e^{H(z_{i})} \right)} \right\rbrack.
\end{equation}

Here, our loss function takes the expected value over the entire training set.
The first term is the mean square error between the input image and the reconstructed image, ensuring image reconstruction performance.
The second term is the product of the length ratio and the number of the activated feature maps. This term represents wireless channel usage and helps reduce the required channel bandwidth. 
The third term indicates the average processed entropy of $z$, which increases the average amount of information carried by each transmitted symbol. It is calculated by taking the expected value of the processed entropy of all the feature maps.
Note that here we do not normalize these entropy values.
We use two hyperparameters $\alpha$ and $\beta$ to balance these three terms.


To train our model, we use the Adam optimizer\cite{kingma2014adam}. 
Specifically, we first train the entire model for 200 epochs with a learning rate of $5 \times 10^{- 4}$. Afterward, we decrease the learning rate to $5 \times 10^{- 5}$ and continue training for another 200 epochs. 
Afterwards, we proceed to the fine-tuning phase, where the learning rate becomes $1 \times 10^{- 5}$.
During this phase, we first freeze the semantic encoder $S_{1}$ and train the model for 100 epochs.
Then, we freeze the semantic encoder $S_{2}$ and the policy networks, and train the model for 100 epochs.


\begin{table*}
\vskip 0.14in
\caption{The strategy comparison between the baseline \cite{yang2022deep} with $\alpha_{\mathrm{base}} = 5 \times 10^{- 4}$ and our method with $\alpha_{\mathrm{ours}} = 2 \times 10^{- 4}$.}
\label{table1}
\centering
\begin{tabular}{c|c|c|c|c|c|c} 
\hline
SNR(dB) & \begin{tabular}[c]{@{}c@{}}Average number\\~of feature maps\end{tabular} & \begin{tabular}[c]{@{}c@{}}Average feature map\\~length ratio(\%)\end{tabular} & Rate(CPP)     & \begin{tabular}[c]{@{}c@{}}Baseline \\Rate(CPP)\end{tabular} & PSNR(dB) & \begin{tabular}[c]{@{}c@{}}Basline \\PSNR(dB)\end{tabular}  \\ 
\hline
15      & 7.90   & 74.04   & \textbf{0.450}    & 0.447 & \textbf{33.37 (↑0.57)}   & 32.80  \\
10      & 7.97   & 100.00  & \textbf{0.498}     & 0.488 & \textbf{31.79 (↑0.59)}  & 31.20  \\
5       & 8.00   & 100.00  & \textbf{0.500}     & 0.499 & \textbf{28.60 (↑0.28)}  & 28.32 \\
0       & 8.00   & 100.00  & \textbf{0.500}     & 0.500 & \textbf{24.74 (↑0.21)}   & 24.53  \\
\hline
\end{tabular}
\end{table*}

\begin{table*}

\caption{
The strategy comparison between the baseline \cite{yang2022deep} with $\alpha_{\mathrm{base}} = 1.5 \times 10^{- 3}$ and our method with $\alpha_{\mathrm{ours}} = 1.3 \times 10^{- 3}$.
}
\label{table2}
\centering
\begin{tabular}{c|c|c|c|c|c|c} 
\hline
SNR(dB) & \begin{tabular}[c]{@{}c@{}}Average number\\~of feature maps\end{tabular} & \begin{tabular}[c]{@{}c@{}}Average feature map\\~length ratio(\%)\end{tabular} & Rate(CPP)      & \begin{tabular}[c]{@{}c@{}}Baseline \\Rate(CPP)\end{tabular} & PSNR(dB)  & \begin{tabular}[c]{@{}c@{}}Basline \\PSNR(dB)\end{tabular}  \\ 
\hline
15      & 4.60  & 73.20   & \textbf{0.260}   & 0.255 & \textbf{30.06(↑0.11)} & 29.95  \\
10      & 5.54  & 100.00  & \textbf{0.346}   & 0.310 & \textbf{29.54(↑0.65)}  & 28.89   \\
5       & 7.29  & 100.00  & \textbf{0.455}   & 0.430 & \textbf{27.88(↑0.40)} & 27.48 \\
0       & 7.90  & 100.00  & \textbf{0.493}   & 0.493 & \textbf{24.56(↑0.11)}  & 24.45 \\
\hline
\end{tabular}
\end{table*}

\begin{figure*}[htbp]

\centering
\subfigure[$\alpha_{\mathrm{ours}} = 2 \times 10^{- 4}$]{
\begin{minipage}[t]{0.48\linewidth}
\label{p1}
\centering
\includegraphics[width=3.15in]{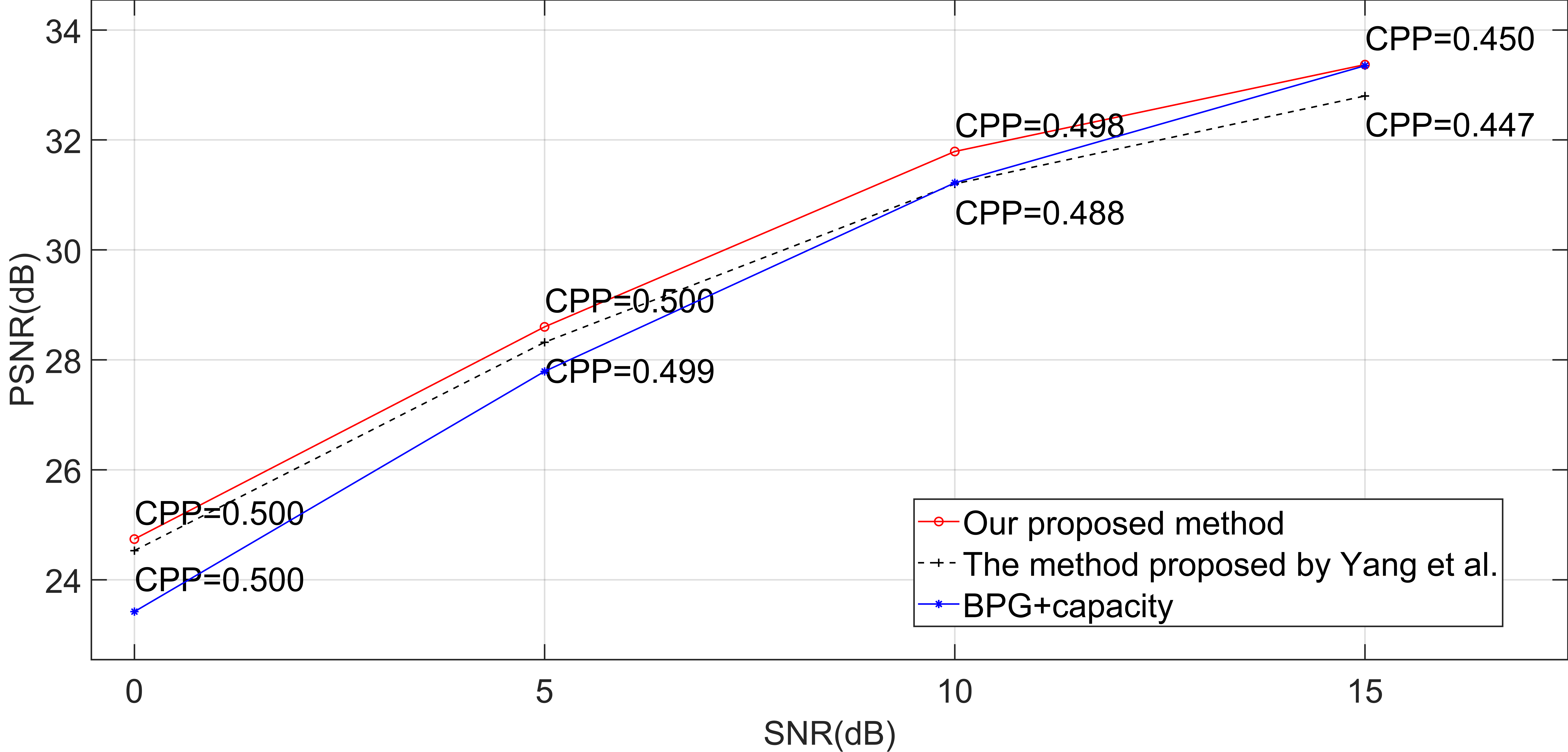}
\end{minipage}
}
\subfigure[$\alpha_{\mathrm{ours}} = 1.3 \times 10^{- 3}$]{
\begin{minipage}[t]{0.48\linewidth}
\label{p1}
\centering
\includegraphics[width=3.15in]{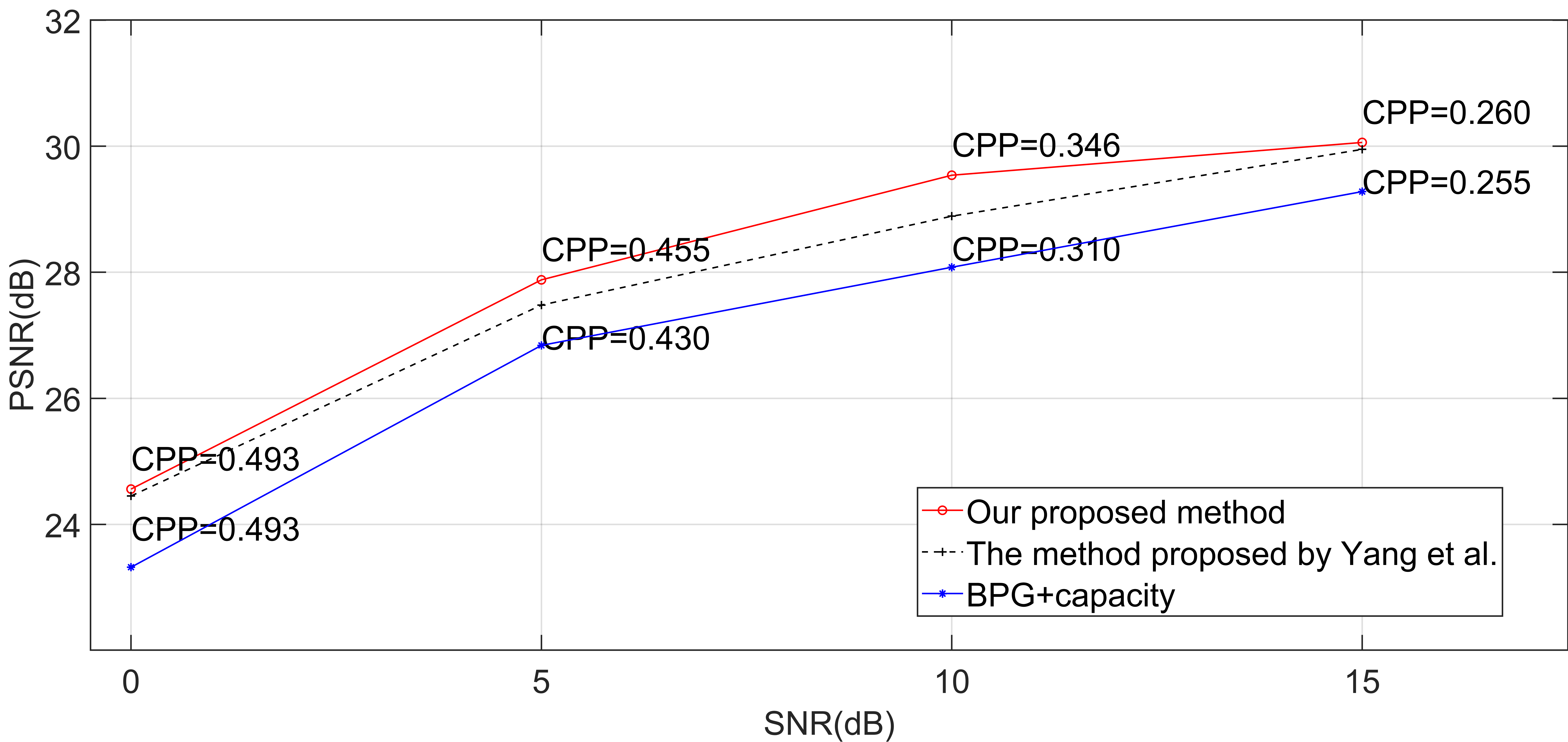}
\end{minipage}
}
\caption{The comparison of image reconstruction performance between our method, the method proposed in \cite{yang2022deep}, and the BPG+capacity method.
}
\label{fig4}
\vskip -0.2in
\end{figure*}

\section{Experiments}
\subsection{Experimental Setup}
We evaluate our proposed scheme using the CIFAR-10 dataset, which consists of 50,000 training images and 10,000 testing images. All images are $32 \times 32$ pixel RGB images.
The batch size is set to 512, and the channel SNR is uniformly sampled between 0dB and 15dB for each image in every epoch during the training process.
The hyperparameter $\alpha_{\mathrm{ours}}$ has two distinct values: $2 \times 10^{- 4}$ and $1.3 \times 10^{- 3}$. 
$\beta$ is set to $1 \times 10^{- 5}$. 
The number of filters in the last layer of the semantic encoder is $2C = 16$, and the length of the corresponding feature maps is $\frac{L}{2} = 64$. 
The feature maps in $z$ are then concatenated in pairs to create selective feature map $z'$ with dimension $C \times L = 8 \times 128$. 
If the feature maps are pruned, the binary vectors used to represent the pruning indices comprise 128 bits.
Thus, $L^{'} = \left\lceil \frac{128}{6} \right\rceil = 22$.

For comparison, we use two baseline methods: the adaptive rate control scheme proposed by Yang et al.\cite{yang2022deep} and the BPG image codec combined with idealistic error-free transmission based on Shannon capacity.
We use peak signal-to-noise ratio (PSNR) as the performance metric.

\subsection{Experimental Results}
First, we analyze the number of feature maps and the pruning ratios selected by our proposed method under different SNRs. 
For comparison, we use data obtained from the baseline \cite{yang2022deep}, which is shown in Table.~\ref{table1} and Table.~\ref{table2}.
To ensure a fair performance comparison, we set $\alpha_{\mathrm{ours}}$ in multiple attempts to maintain CPP levels similar.
When the SNR is low (0-10dB), our method selects a larger number of feature maps and we do not prune them in this case.
This is because poor channel conditions require more feature maps to be transmitted to provide enough information for image reconstruction.
Pruning is not worthwhile in this case because 64-QAM has a high bit error rate (BER) at low SNRs. 
Our method achieves approximately 0.11-0.28dB higher PSNR compared to the baseline without pruning the feature maps.
This indicates that our designed entropy-aware feature map selection module can help improve image reconstruction performance.
As the SNR increases to 15dB, we prune about 26\%-27\% of the pixels in the activated feature maps.

Next, we compare our proposed method with the method proposed by Yang et al. \cite{yang2022deep} and the BPG+capacity method. 
We compare the performance of these models and plot the SNR-PSNR curves, as shown in Fig.~\ref{fig4}.
Our model outperforms the method proposed in \cite{yang2022deep} for both $\alpha_{\mathrm{ours}} = 2 \times 10^{- 4}$ and $\alpha_{\mathrm{ours}} = 1.3 \times 10^{- 3}$. 
At similar CPPs, our method improves PSNR by about 0.21-0.57dB when $\alpha_{\mathrm{ours}}$ is relatively small.
However, the PSNR improvement drops to around 0.11dB when we increase the value of $\alpha_{\mathrm{ours}}$. 
This is because that as $\alpha_{\mathrm{ours}}$ increases, the policy networks focus more on channel bandwidth usage rather than PSNR performance, thereby reducing the effectiveness of our scheme.
At low SNRs (0-10dB), our method outperforms both baseline methods.
At high SNR (15dB), our method performs equivalently to the BPG+capacity method when $\alpha_{\mathrm{ours}}$ is small, while completely outperforming it when $\alpha_{\mathrm{ours}}$ increases. Note that the BPG+capacity method is rate-fixed. Therefore, even if the PSNR performances are similar, our method has the advantage of rate adaptivity.
Compared to the method proposed in \cite{yang2022deep}, our method shows a notable improvement in image reconstruction performance, achieving approximately 0.57dB for small $\alpha_{\mathrm{ours}}$ and slightly lower, about 0.11dB, for larger $\alpha_{\mathrm{ours}}$. 
This shows that our method is more advantageous when channel resources are sufficient.

\begin{figure}[ht]
\begin{center}
\centerline{\includegraphics[width=0.75\linewidth]{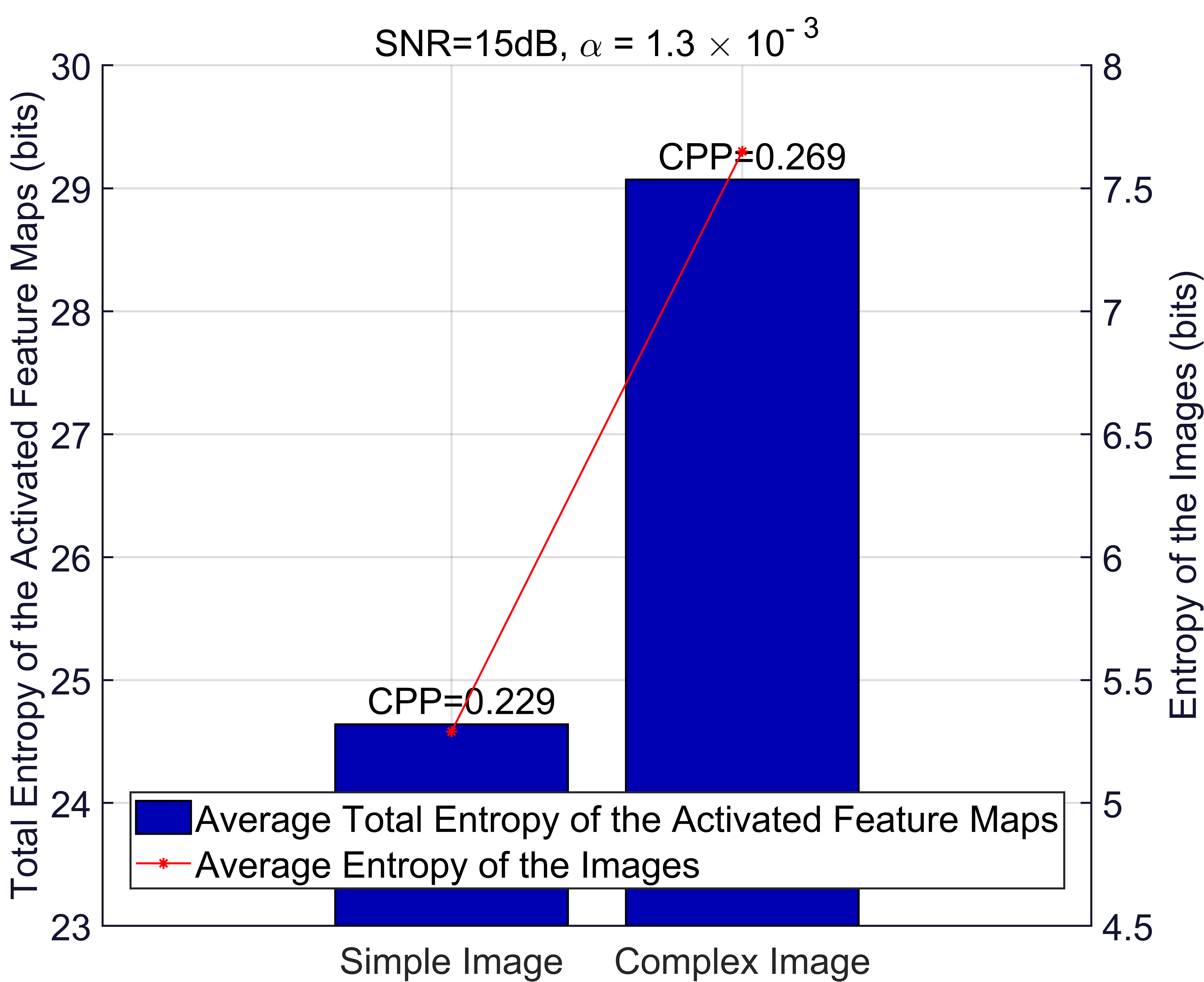}}
\caption{The comparison of transmission schemes for two types of images with different complexity in CIFAR-10.}
\label{fig5}
\end{center}
\vskip -0.1in
\end{figure}

\begin{table}
\caption{The performance comparison of the proposed method with and without the feature map pruning technique.}
\label{table3}
\centering
\begin{tabular}{c|c|c} 
\hline
Model        & Rate(CPP)     & PSNR(dB)      \\ 
\hline
$\alpha_{\mathrm{ours}} = 2 \times 10^{- 4}$ (with $P_{2}$)    & 0.450  & 33.37(↑0.18)         \\
$\alpha_{\mathrm{ours}} = 4.5 \times 10^{- 4}$ (without $P_{2}$) & 0.451         & 33.19         \\
\hline

$\alpha_{\mathrm{ours}} = 3.5 \times 10^{- 4}$ (with $P_{2}$)    & 0.433  & 33.10(↑0.16)         \\
$\alpha_{\mathrm{ours}} = 4.8 \times 10^{- 4}$ (without $P_{2}$) & 0.433  & 32.94         \\
\hline

$\alpha_{\mathrm{ours}} = 4 \times 10^{- 4}$ (with $P_{2}$)    & 0.406  & 32.65(↑0.09)         \\
$\alpha_{\mathrm{ours}} = 5.4  \times 10^{- 4}$ (without $P_{2}$) & 0.405  & 32.56         \\
\hline

$\alpha_{\mathrm{ours}} = 5.8 \times 10^{- 4}$ (with $P_{2}$)    & 0.356  & 31.77(↑0.02)         \\
$\alpha_{\mathrm{ours}} = 7.5 \times 10^{- 4}$ (without $P_{2}$) & 0.356         & 31.75         \\
\hline
$\alpha_{\mathrm{ours}} = 1.3 \times 10^{- 3}$ (with $P_{2}$)    & 0.260         & 30.06(↑0.02)  \\
$\alpha_{\mathrm{ours}} = 1.9 \times 10^{- 3}$ (without $P_{2}$) & 0.260         & 30.04         \\
\hline
\end{tabular}
\vskip -0.1in
\end{table}





Finally, we consider two types of images in CIFAR-10 transmitted by our proposed scheme, as shown in Fig.~\ref{fig5}. 
In Fig.~\ref{fig5}, the blue bar represents the total entropy of the activated feature maps, while the red curve represents the entropy of the input images, both of which refer to the average.
The two types of images have different complexity, with higher entropy representing higher complexity. 
For each type of image, we take the average of 100 images.
For the simple images, our method selects an average of 4.11 feature maps and prunes 28.2\% pixels of them, resulting in an average CPP of 0.229. While for the complex images, our method selects an average of 4.83 feature maps and prunes 27.9\% pixels of them, resulting in an average CPP of 0.269. 
The result shows that our method selects fewer feature maps and therefore lower CPP when the image content is relatively simple, and vice versa. It also proves that our method can automatically select activated feature maps based on the entropy of the image.
In conclusion, by combining the entropy-aware feature map selection with the feature map pruning, we achieve notable improvement in image reconstruction performance.


\subsection{Ablation Study}
We also conduct an ablation study of the proposed scheme to evaluate its effectiveness, as shown in Table.~\ref{table3}.
Specifically, we train several models with different $\alpha_{\mathrm{ours}}$ values to compare the performance of models with and without the feature map pruning technique.
To ensure a fair performance comparison, we set the $\alpha_{\mathrm{ours}}$ values through multiple attempts to ensure similar CPP levels. 

By comparing the results, we find that pruning the activated feature maps can improve image reconstruction performance of the model with similar CPPs. 
The effect of the feature map pruning technique is more significant at lower $\alpha_{\mathrm{ours}}$.
These experiments confirm that our feature map pruning technique effectively improves the image reconstruction performance.


\section{Conclusion}
In this paper, we proposed a deep JSCC scheme for wireless image transmission with entropy-aware adaptive rate control.
During training, we maximized the entropy of the feature maps to increase the amount of average information carried by each transmitted symbol, and selected activated feature maps based on their entropy.
This helped the model to achieve better image reconstruction performance. 
To achieve the adaptive transmission rate, we introduced two policy networks. 
The first generated a binary mask that selects activated feature maps. 
Subsequently, the second policy network decided the pruning ratio for the activated feature maps.
We transmitted the position information of the pruned pixels over the wireless channel to recover the structure of the activated feature maps at the receiver.
Our experiments demonstrate that the proposed method learns an effective rate control strategy that automatically adjusts the rate according to the feature maps and their entropy, as well as the channel conditions. 
Our method, which benefits from the entropy-aware feature map selection and the feature map pruning, provides a promising experimental result for adaptive rate control in deep JSCC. 
Moreover, it outperforms existing related studies in terms of image reconstruction performance.

\vspace{12pt}
\end{CJK}

\begin{thebibliography}{10}
\providecommand{\url}[1]{#1}
\csname url@samestyle\endcsname
\providecommand{\newblock}{\relax}
\providecommand{\bibinfo}[2]{#2}
\providecommand{\BIBentrySTDinterwordspacing}{\spaceskip=0pt\relax}
\providecommand{\BIBentryALTinterwordstretchfactor}{4}
\providecommand{\BIBentryALTinterwordspacing}{\spaceskip=\fontdimen2\font plus
\BIBentryALTinterwordstretchfactor\fontdimen3\font minus
  \fontdimen4\font\relax}
\providecommand{\BIBforeignlanguage}[2]{{%
\expandafter\ifx\csname l@#1\endcsname\relax
\typeout{** WARNING: IEEEtran.bst: No hyphenation pattern has been}%
\typeout{** loaded for the language `#1'. Using the pattern for}%
\typeout{** the default language instead.}%
\else
\language=\csname l@#1\endcsname
\fi
#2}}
\providecommand{\BIBdecl}{\relax}
\BIBdecl
\bibitem{cover1999elements}
T.~M. Cover, \emph{Elements of information theory}.\hskip 1em plus 0.5em minus
  0.4em\relax John Wiley \& Sons, 1999.
  






\bibitem{bourtsoulatze2019deep}
E.~Bourtsoulatze, D.~B. Kurka, and D.~G{\"u}nd{\"u}z, ``Deep joint
  source-channel coding for wireless image transmission,'' \emph{IEEE
  Transactions on Cognitive Communications and Networking}, vol.~5, no.~3, pp.
  567--579, 2019.


  



\bibitem{kurka2020deep}
D.~B. Kurka and D.~G{\"u}nd{\"u}z, ``Deep joint source-channel coding of images
  with feedback,'' in \emph{ICASSP 2020-2020 IEEE International Conference on
  Acoustics, Speech and Signal Processing (ICASSP)}.\hskip 1em plus 0.5em minus
  0.4em\relax IEEE, 2020, pp. 5235--5239.

\bibitem{zhang2022wireless}
Z.~Zhang, Q.~Yang, S.~He, M.~Sun, and J.~Chen, ``Wireless transmission of
  images with the assistance of multi-level semantic information,'' in
  \emph{2022 International Symposium on Wireless Communication Systems
  (ISWCS)}.\hskip 1em plus 0.5em minus 0.4em\relax IEEE, 2022, pp. 1--6.
 \bibitem{kurka2019successive}
D.~B. Kurka and D.~G{\"u}nd{\"u}z, ``Successive refinement of images with deep
  joint source-channel coding,'' in \emph{2019 IEEE 20th International Workshop
  on Signal Processing Advances in Wireless Communications (SPAWC)}.\hskip 1em
  plus 0.5em minus 0.4em\relax IEEE, 2019, pp. 1--5.

\bibitem{kurka2021bandwidth}
D.~B. Kurka and D.~G{\"u}nd{\"u}z, ``Bandwidth-agile image transmission with deep joint source-channel
  coding,'' \emph{IEEE Transactions on Wireless Communications}, vol.~20,
  no.~12, pp. 8081--8095, 2021.
  
\bibitem{yang2022deep}
M.~Yang and H.-S. Kim, ``Deep joint source-channel coding for wireless image
  transmission with adaptive rate control,'' in \emph{ICASSP 2022-2022 IEEE
  International Conference on Acoustics, Speech and Signal Processing
  (ICASSP)}.\hskip 1em plus 0.5em minus 0.4em\relax IEEE, 2022, pp. 5193--5197.

\bibitem{zhou2022adaptive}
Q.~Zhou, R.~Li, Z.~Zhao, Y.~Xiao, and H.~Zhang, ``Adaptive bit rate control in
  semantic communication with incremental knowledge-based harq,'' \emph{IEEE
  Open Journal of the Communications Society}, vol.~3, pp. 1076--1089, 2022.

\bibitem{han2015learning}
S.~Han, J.~Pool, J.~Tran, and W.~Dally, ``Learning both weights and connections
  for efficient neural network,'' \emph{Advances in neural information
  processing systems}, vol.~28, 2015.
  






\bibitem{kingma2014adam}
D.~P. Kingma and J.~Ba, ``Adam: A method for stochastic optimization,''
  \emph{arXiv preprint arXiv:1412.6980}, 2014.



\end{thebibliography}
\end{document}